\documentstyle[12pt]{article}
\makeatletter
\@addtoreset{equation}{section}

\makeatother

\begin{document}

\begin{flushright}
Jun 2001

KEK-TH-771
\end{flushright}

\begin{center}

\vspace{5cm}

{\Large Closed String Tachyons in} 

\vspace{3mm}

{\Large Non-supersymmetric Heterotic Theories}

\vspace{2cm}

Takao Suyama \footnote{e-mail address : tsuyama@post.kek.jp}

\vspace{1cm}

{\it Theory Group, KEK}

{\it Tsukuba, Ibaraki 305-0801, Japan}

\vspace{4cm}

{\bf Abstract} 

\end{center}

We calculate partition functions for supersymmetric heterotic theories on Melvin background 
with Wilson line. 
These functions coincide with partition functions for some of non-supersymmetric heterotic 
theories in appropriate limits. 
This suggests that non-supersymmetric heterotic theories are equivalent to supersymmetric 
theories on supersymmetry-breaking backgrounds, just as in the case of recently conjectured 
IIA-0A duality.

\newpage

\vspace{1cm}

\section{Introduction}

\vspace{5mm}

String theory has infinitely many perturbative vacua. 
There are many 
non-supersymmetric vacua which in general contain tachyons in their spectrum, as well as 
supersymmetric ones. 
The existence of tachyons indicates an instability of the perturbative vacuum, and it has 
been believed that the theory becomes stable after condensing the tachyon. 
In recent researches \cite{Open}, the fate of an open string tachyon has been well-studied. 
The tachyon has a nontrivial potential and the stable vacuum of the theory corresponds to 
the global minimum of the potential. 
The stabilization corresponds to the decay of an unstable D-brane system on which the open 
string is attached. 
The occurrence of this phenomenon has been verified quantitatively. 

The case of closed string tachyon is rather difficult to analyze. 
Recently, an interesting conjecture is made in \cite{Melvin}\cite{fluxbrane} on the fate of 
closed string tachyon in Type 0A theory (see also \cite{RT}). 
It is argued that Type 0A theory is equivalent to Type IIA theory on Melvin background 
\cite{MelvinSoln} with a particular value of parameters. 
Then the instability due to the closed string tachyon in Type 0A theory is identified with 
the one due to 
pair creation of D6-$\overline{\mbox{D6}}$, and it is argued that the theory becomes Type IIA 
theory after closed string tachyon condensation \cite{Melvin}\cite{fluxbrane}. 
This picture seems to be correct, so it is natural to expect that such understanding can be 
applicable to other theories containing closed string tachyons. 

In this paper, we consider various non-supersymmetric heterotic theories \cite{nonhet}. 
All such theories but one have a closed string tachyon, so these are good places to examine 
the range of applicability of the above picture. 

Type IIA-Type 0A duality is based on the following facts. 
Type 0A theory can be obtained by twisting Type IIA by $(-1)^{F_s}$, where $F_s$ is the 
spacetime fermion number. 
Moreover there is a Type IIA orbifold which interpolates between Type IIA and Type 0A 
\cite{orbifold}\cite{Type0}. 
The corresponding investigations in heterotic theory have already done \cite{hetorb}. 
And in \cite{Melvin}, it is shown that the partition function of the orbifold coincides 
with that of Type IIA theory on the Melvin background. 
Therefore it is natural to expect that there are some backgrounds of heterotic theory which 
are dual to non-supersymmetric heterotic theories. 

This paper is organized as follows. 
The Melvin background and its relation to Type 0A theory is reviewed in section 2. 
The main piece of evidence of IIA-0A duality is briefly shown in section 3. 
In section 4, we calculate partition functions for supersymmetric heterotic theories on 
the Melvin background with several patterns of Wilson lines, and show that the partition 
functions coincide with those for some of non-supersymmetric heterotic theories. 
Section 5 is devoted to discussions.

\vspace{1cm}

\section{Kaluza-Klein Melvin Background}

\vspace{5mm}

Consider the following spacetime in M-theory \cite{Melvin},
\begin{equation}
ds_{11}^2 = \eta_{\mu\nu}dx^{\mu}dx^{\nu}+dr^2+r^2(d\theta+qdy)^2+dy^2,
  \label{11dim}
\end{equation}
with $\mu,\nu=0,1,\cdots,7$. 
$(r,\theta)$ are polar coordinates of 8-9 plane and $y$-direction is compactified on $S^1$. 
That is, 
\begin{eqnarray}
&& y \sim y+2\pi mR, \nonumber \\
&& \theta \sim \theta+2\pi n, 
  \label{identify} \\
&& \hspace*{5mm} m,n\in {\bf Z}, \nonumber
\end{eqnarray}
where $R$ is the radius of the $S^1$ and $q$ is a real parameter. 
This metric is locally flat, so it is expected that this is an exact background of M-theory. 

Although the eleven-dimensional metric is simple, the corresponding Type IIA background is 
nontrivial. 
The ten-dimensional metric and other fields in Type IIA theory can be read off from the 
metric (\ref{11dim}) by the well-known relation,
\begin{equation}
ds_{11}^2 = e^{-\frac23\phi}ds_{10}^2+e^{\frac43\phi}(dy+A_\mu dx^{\mu})^2.
\end{equation}
The followings are the corresponding Type IIA background known as the Kaluza-Klein Melvin 
background \cite{Melvin}. 
\begin{eqnarray}
ds_{10}^2 &=& \sqrt{1+q^2r^2}\ (\eta_{\mu\nu}dx^{\mu}dx^{\nu}+dr^2)
             +\frac{r^2}{\sqrt{1+q^2r^2}}d\theta^2 \nonumber \\
e^{\frac43\phi} &=& 1+q^2r^2 
  \label{10dim} \\
A_\theta &=& \frac{qr^2}{1+q^2r^2} \nonumber 
\end{eqnarray}
Since the dilaton diverges as $r\to \infty$, this is not a perturbative background. 
The only nonzero component of the R-R field strength is 
\begin{equation}
F_{89}=\frac{2q}{(1+q^2r^2)}\ . 
\end{equation}
This indicates that the R-R flux is localized around $r=0$. 
The total flux is finite: 
\begin{equation}
\frac1{2\pi}\int_{{\bf R}^2} F = \frac1q\ .
\end{equation}
This configuration is called a fluxbrane or a F7-brane \cite{fluxbrane}. 
Further investigations on the fluxbranes can be found in \cite{fluxbrane2} 

\vspace{2mm}

Surprisingly, there is a periodicity in $q$ \cite{Melvin}. 
This periodicity is rather apparent in the eleven-dimensional solution, although it seems 
difficult to find such a property from the ten-dimensional one. 
It is convenient to introduce new coordinate $\tilde{\theta}=\theta+qy$. 
Then the identifications of the coordinates (\ref{identify}) become as follows.
\begin{eqnarray}
&& y \sim y+2\pi mR \nonumber \\
&& \tilde{\theta} \sim \tilde{\theta}+2\pi n+2\pi mqR \nonumber 
\end{eqnarray}
One can see that when $q=\frac kR$ ($k$ is an integer), $\tilde{\theta}$ is just the 
ordinary angular variable. 
Therefore the corresponding spacetime is ${\bf R}^{10}\times S^1$ which is the same as the 
one with $q=0$. 

\vspace{2mm}

If there exist spacetime fermions, the situation is a bit complicated. 
Suppose that there is a fermion whose boundary condition is periodic along the $S^1$. 
When $q=\frac1R$, the parallel transport of the fermion along the $S^1$ is accompanied by 
the $2\pi$ rotation in the 8-9 plane. 
So the boundary condition is antiperiodic at this special value of $q$. 
When $q=\frac2R$, the boundary condition is again periodic because of the $4\pi$ rotation in 
the 8-9 plane, and thus this background is indeed equivalent to the one with $q=0$. 
Therefore the period of $q$, in the presence of fermions, is $\frac2R$. 

The background with $q=\frac1R$, when compactified to ten dimensions, corresponds to Type 0A 
theory \cite{Type0}. 
Type 0A theory is a theory of closed strings which contain a tachyonic mode. 
The existence of such a closed string tachyon would indicate an instability of the vacuum, 
just as in the case of open string tachyon \cite{Open}. 
The correspondence discussed above means that Type 0A theory can be interpreted as Type 
IIA theory on the particular background. 
Based on this point of view, it is argued that the instability due to the closed string 
tachyon in Type 0A theory corresponds, in Type IIA theory, to the one due to pair creation 
of D6-branes \cite{Melvin}\cite{fluxbrane} (instability of the Melvin background is 
investigated in \cite{instability}). 
After the closed string tachyon is stabilized in some manner, Type 0A theory would 
become Type IIA theory.

\vspace{1cm}

\section{Duality between Type IIA and Type 0A}

\vspace{5mm}

The main piece of evidence of the duality between Type IIA and Type 0A is the coincidence 
of their partition functions on appropriate backgrounds \cite{Melvin}. 
As mentioned before, perturbative calculations are not valid in the background 
(\ref{10dim}). 
In addition, it is very difficult to analyze string theory on R-R background. 
Fortunately, the background (\ref{10dim}) can be related to a NS-NS background by further 
compactifying the metric (\ref{11dim}) on $S^1$ along, say, 7-direction and reducing to ten 
dimensions along the $S^1$. 
The resulting Type IIA background is as follows. 
\begin{equation}
ds_{NS}^2 = \eta_{\alpha\beta}dx^{\alpha}dx^{\beta}+dr^2+r^2(d\theta+qdy)^2+dy^2
  \label{NS}
\end{equation}
The indices $\alpha, \beta$ run from 0 to 6. 
The dilaton is constant and there is no R-R background. 

It is shown in \cite{solve} that Type IIA theory on the background (\ref{NS}) can be reduced 
to a free theory, so the partition function can be calculated exactly. 
Note that since the background (\ref{NS}) is locally flat, the worldsheet theory is 
conformal for any value of $q$ \cite{conformal}. 

The partition function for general $q$ is \cite{solve}\cite{Melvin}
\begin{eqnarray}
Z_q &\propto& \int \frac{d^2\tau}{\tau_2}\tau^{-4}|\eta(\tau)|^{-18}\sum_{m,n\in {\bf Z}}
           \exp\left(-\frac{\pi R^2}{\alpha'\tau_2}|n+m\tau|^2\right) \nonumber \\
    & & \left|\vartheta {1/2+mqR \brack 1/2+nqR}(0,\tau)\right|^{-2}
              \left|\vartheta{1/2+mqR/2 \brack 1/2+nqR/2}(0,\tau)\right|^8. 
\end{eqnarray}
We have used the theta function with characteristics
\begin{equation}
\vartheta{a \brack b}(\nu,\tau) 
   = \sum_{n\in {\bf Z}}\exp\left[\pi i(n+a)^2\tau+2\pi i(n+a)(\nu+b)\right].
\end{equation}
The details of this calculation will be explained in the next section. 
From the above expression, one can see the periodicity in $q$ \cite{solve},
\begin{equation}
Z_{q+2/R} = Z_q,
\end{equation}
discussed in the previous section. 

Since Type IIA theory contains spacetime fermions, the shift of $q$ by $\frac1R$ is expected 
to make a change. 
In fact, the partition function becomes
\begin{eqnarray}
Z_{1/R} &\propto& \int \frac{d^2\tau}{\tau_2}\tau_2^{-5}|\eta(\tau)|^{-24} \nonumber \\
        & &  \left( Z_{(1,0)}\left|\vartheta{1 \brack 0}(0,\tau)\right|^8 
                   +Z_{(0,1)}\left|\vartheta{0 \brack 1}(0,\tau)\right|^8 
                   +Z_{(1,1)}\left|\vartheta{0 \brack 0}(0,\tau)\right|^8\right)\ ,
\end{eqnarray}
where 
\begin{equation}
Z_{(\epsilon,\delta)} = \sum_{k,l\in {\bf Z}}\exp\left[-\frac{\pi R^2}{\alpha'\tau_2}
                         |(2k+\epsilon)+(2l+\delta)\tau|^2\right].
\end{equation}
In the $R\to 0$ limit, 
\begin{equation}
Z_{1/R}\to \int \frac{d^2\tau}{\tau_2}\tau_2^{-5}|\eta(\tau)|^{-16}
             \left( |Z^1_0(\tau)|^8+|Z^0_1(\tau)|^8+|Z^0_0(\tau)|^8 \right)\ ,
  \label{dual}
\end{equation}
where $Z^\epsilon_\delta(\tau)$ is the partition function for a complex fermion defined in 
\cite{Pol}. 
The RHS of (\ref{dual}) is the partition function for Type 0A theory compactified on 
$S^1$ with vanishing radius. 
In the eleven-dimensional background (\ref{11dim}), the $R\to 0$ limit just corresponds to 
Type IIA limit. 
Therefore (\ref{dual}) can be understood as evidence of the duality between 
Type IIA on the Melvin background and Type 0A. 
It can also be interpreted that (\ref{dual}) shows a duality between 
ten-dimensional theories in the T-dual picture.

\vspace{1cm}

\section{Non-supersymmetric heterotic theory}

\vspace{5mm}

So far we have reviewed the duality between Type IIA theory on the Melvin background and 
Type 0A theory. 
This duality leads one to the conjecture on the fate of the closed string tachyon in Type 0A 
theory \cite{Melvin}\cite{fluxbrane}. 
Another approach to closed string tachyon has been made recently \cite{AdS}. 
The stabilization of closed string tachyon is a phenomenon which is difficult to study, 
although that of open string tachyon has been well-studied \cite{Open}. 
Therefore it will be very helpful for understanding the fate of closed string tachyons if 
such a dual picture exists in general. 

It is well-known that there are several non-supersymmetric heterotic theories in ten 
dimensions \cite{nonhet} and all the theories but one have tachyons in their spectrum. 
Thus it is natural to expect that these tachyonic theories can also be interpreted as 
supersymmetric heterotic theory on particular backgrounds which break all of 
supersymmetries. 
In this section we will show evidence which suggests the existence of such dual 
descriptions.

\vspace{5mm}

\subsection{Background of heterotic theory}

\vspace{2mm}

We consider the NS-NS Melvin background (\ref{NS}) and a Wilson line along the 
$y$-direction. 
It is convenient to rewrite the metric as follows. 
\begin{equation}
ds_{NS}^2 = \eta_{\mu\nu}dx^{\mu}dx^{\nu}+|dx+iqxdy|^2+dy^2
\end{equation}
We have introduced a complex coordinate $x$,
\begin{equation}
x=re^{i\theta}=x^7+ix^8.
\end{equation}
Then the bosonic part of the worldsheet action of heterotic theory on this background 
is \cite{solve}
\begin{equation}
S_b = \frac1{4\pi\alpha'}\int d^2\sigma\left[\eta_{\mu\nu}\partial_aX^\mu\partial_aX^\nu
       +|\partial_aX+iq\partial_ayX |^2+\partial_ay\partial_ay\right], 
\end{equation}
For the right-moving fermions, we employ the Green-Schwarz formalism. 
Their action is \cite{solve}
\begin{equation}
S_{right} = \frac i\pi\int d^2\sigma \bar{S}^r_R\left(\partial_++\frac i2q\partial_+y\right)
S^r_R\ .
\end{equation}
with $r=1,\cdots,4$. 
Note that the coefficient of $\partial_+y$ is half of that in the bosonic part. 

The action of the left-moving fermions is not affected by the Melvin background. 
Instead, by the presence of the Wilson line the action is as follows. 
\begin{equation}
S_{left} = \frac i\pi \int d^2\sigma \sum_{I=1}^{16}
       \bar{\lambda}_L^I(\partial_--iA_y^I\partial_-y)\lambda_L^I
\end{equation}
As mentioned in the previous section, this theory is conformal for any value of $q$ 
(and $A_y^I$). 

\vspace{2mm}

At first sight, the above action is too complicated to analyze, due to the nontrivial 
coupling between $y$ and other fields. 
However, as shown in \cite{solve}, this can be reduced to a free theory. 
The reduction is accomplished by the following field redefinitions. 
\begin{eqnarray}
&& X = e^{-iqy}\tilde{X} \\
&& S_R^r = e^{-\frac i2qy}\tilde{S}_R^r  \\
&& \lambda_L^I = e^{iA_y^Iy}\tilde{\lambda}_L^I  
\end{eqnarray}
The Jacobian of this redefinition is trivial. 
The action of $\tilde{X}$, $\tilde{S}^r_R$ and $\tilde{\lambda}^I_L$ becomes the free one. 
The existence of the nontrivial background is encoded in the boundary conditions. 
\begin{eqnarray}
&& y(\sigma+2\pi) = y(\sigma)+2\pi mR \\
&& \tilde{X}(\sigma+2\pi) = e^{2\pi imqR}\tilde{X}(\sigma) \\
&& \tilde{S}_R^r(\sigma+2\pi) = e^{\pi imqR}\tilde{S}_R^r(\sigma) \\
&& \tilde{\lambda}_L^I(\sigma+2\pi) = e^{-2\pi imRA_y^I}\tilde{\lambda}_L^I(\sigma)
\end{eqnarray}

The torus partition function of this theory can be easily calculated, for example, by the 
operator formalism. 
For a complex boson with boundary conditions 
\begin{eqnarray}
&& \phi(\sigma^1+2\pi,\sigma^2) = e^{2\pi i\theta_1}\phi(\sigma^1,\sigma^2), \\
&& \phi(\sigma^1+2\pi \tau_1,\sigma^2+2\pi \tau_2) 
     = e^{-2\pi i\theta_2}\phi(\sigma^1,\sigma^2), 
\end{eqnarray}
the partition function is $Z^{1,2\theta_1}_{1,2\theta_2}(\tau)^{-1}$, where
\begin{equation}
Z^{1,2\theta_1}_{1,2\theta_2}(\tau)
      \propto \vartheta{1/2+\theta_1 \brack 1/2+\theta_2}(0,\tau)\eta(\tau)^{-1}
\end{equation}
and $\tau$ is the modulus of the torus. 
Similarly, for a complex fermion with boundary conditions
\begin{eqnarray}
&& \psi(\sigma^1+2\pi,\sigma^2) = -e^{2\pi i\theta_1}\psi(\sigma^1,\sigma^2), \\
&& \psi(\sigma^1+2\pi \tau_1,\sigma^2+2\pi \tau_2) 
     = -e^{-2\pi i\theta_2}\psi(\sigma^1,\sigma^2), 
\end{eqnarray}
the partition function is $Z^{0,2\theta_1}_{0,2\theta_2}(\tau)$. 

In the case of heterotic theories, the phases of the partition functions of the fermions 
are important. 
They are determined in \cite{phase} by an argument of anomaly. 
The precise expression, including the phases, of the partition functions is as follows. 
\begin{eqnarray}
&& Z^{0,2\theta_1}_{0,2\theta_2}(\tau) = e^{-\pi i\theta_1\theta_2}
        \vartheta{\theta_1 \brack \theta_2}(0,\tau)\eta(\tau)^{-1}
        \hspace{3.1cm} \mbox{(AA)} \nonumber \\
&& Z^{0,2\theta_1}_{1,2\theta_2}(\tau) = e^{-\pi i\theta_1(\theta_2+1)}
        \vartheta{\theta_1 \brack 1/2+\theta_2}(0,\tau)\eta(\tau)^{-1}
        \hspace{1cm} \mbox{(AP)} \nonumber \\
&& Z^{1,2\theta_1}_{0,2\theta_2}(\tau) = e^{-\pi i\theta_1\theta_2}
        \vartheta{1/2+\theta_1 \brack \theta_2}(0,\tau)\eta(\tau)^{-1}
        \hspace{1.6cm} \mbox{(PA)} \nonumber \\
&& Z^{1,2\theta_1}_{1,2\theta_2}(\tau) = -ie^{-\pi i\theta_1(\theta_2+1)}
        \vartheta{1/2+\theta_1 \brack 1/2+\theta_2}(0,\tau)\eta(\tau)^{-1}
        \hspace{9mm} \mbox{(PP)}
  \label{PFuns}
\end{eqnarray}
Here P (A) indicates (anti-)periodic boundary condition, so, for example, (AP) means 
antiperiodic along $\sigma^1$ direction and periodic along $\sigma^2$ direction. 
The modular property of these functions is summarized in Appendix A. 

\vspace{2mm}

Now one can obtain the partition function for supersymmetric heterotic theories on the 
NS-NS Melvin background with the Wilson line, 
\begin{eqnarray}
&& Z_q(A_y^I) \nonumber \\
&=& \int \frac{d^2\tau}{\tau_2}\tau_2^{-4}|\eta(\tau)|^{-12}
          \sum_{m,n\in {\bf Z}}\exp\left[-\frac{\pi R^2}{\alpha'\tau_2}|n+m\tau|^2\right]
          \left|Z^{1,2mqR}_{1,2nqR}(\tau)\right|^{-2}Z_{left}(\tau)
          {Z^{1,mqR}_{1,nqR}(\tau)^*}^4, \nonumber \\
  \label{PartFn}
\end{eqnarray}
where
\begin{equation}
Z_{left}(\tau) = \prod_{I=1}^{16}Z^{0,2mRA_y^I}_{0,2nRA_y^I}(\tau)
                +\prod_{I=1}^{16}Z^{0,2mRA_y^I}_{1,2nRA_y^I}(\tau)
                +\prod_{I=1}^{16}Z^{1,2mRA_y^I}_{0,2nRA_y^I}(\tau)
                +\prod_{I=1}^{16}Z^{1,2mRA_y^I}_{1,2nRA_y^I}(\tau) 
\end{equation}
for the $SO(32)$ theory, and 
\begin{eqnarray}
 Z_{left}(\tau) &=& \prod_{k=0,1}\left(
                 \prod_{I=1+8k}^{8+8k}Z^{0,2mRA_y^I}_{0,2nRA_y^I}(\tau)
                +\prod_{I=1+8k}^{8+8k}Z^{0,2mRA_y^I}_{1,2nRA_y^I}(\tau)
                   \right. \nonumber \\
&& \hspace*{1cm} \left.
                +\prod_{I=1+8k}^{8+8k}Z^{1,2mRA_y^I}_{0,2nRA_y^I}(\tau)
                +\prod_{I=1+8k}^{8+8k}Z^{1,2mRA_y^I}_{1,2nRA_y^I}(\tau) 
                 \right)^2
\end{eqnarray}
for the $E_8\times E_8$ theory.

\vspace{5mm}

\subsection{Coincidence between partition functions}

\vspace{2mm}

We evaluate the above partition function (\ref{PartFn}) at $q=\frac1R$ in the 
$R\to 0$ limit with the following Wilson lines; 

\vspace{1mm}

\hspace*{5mm}(i) $ A_y^I = (1,0^{15})/R$ ,

\vspace{1mm}

\hspace*{5mm}(ii) $A_y^I = (1,0^7,1,0^7)/R$ \ in $E_8\times E_8$ theory and 

\hspace*{12mm}$ A_y^I = \left(\left(\frac12\right)^8,0^8\right)/R$ \ in $SO(32)$ theory,

\vspace{1mm}

\hspace*{5mm}(iii) no Wilson line.

\vspace{1cm}

(i) Non-SUSY theory containing tachyons

\vspace{2mm}

First we consider $SO(32)$ heterotic theory. 
Then the partition function (\ref{PartFn}) is 
\begin{eqnarray}
Z_{1/R}(1,0^{15}) &=& \int \frac{d^2\tau}{\tau_2}\tau_2^{-5}|\eta(\tau)|^{-16}
          \sum_{m,n\in {\bf Z}}\exp\left[-\frac{\pi R^2}{\alpha'\tau_2}|n+m\tau|^2\right]
          {Z^{1,m}_{1,n}(\tau)^*}^4
           \nonumber \\
   & & \hspace*{5mm} \left( Z^{0,2m}_{0,2n}(\tau)Z^0_0(\tau)^{15}
             +Z^{0,2m}_{1,2n}(\tau)Z^0_1(\tau)^{15}
             +Z^{1,2m}_{0,2n}(\tau)Z^1_0(\tau)^{15} \right) \nonumber \\
   &=& \int \frac{d^2\tau}{\tau_2}\tau_2^{-5}|\eta(\tau)|^{-16}
       \left\{ Z_{(1,0)}\left( Z^0_0(\tau)^{16}+Z^0_1(\tau)^{16}-Z^1_0(\tau)^{16} \right)
              {Z^1_0(\tau)^*}^4 \right. \nonumber \\
   & & \hspace*{3cm} 
        +Z_{(0,1)}\left( Z^0_0(\tau)^{16}-Z^0_1(\tau)^{16}+Z^1_0(\tau)^{16} \right)
              {Z^0_1(\tau)^*}^4 \nonumber \\
   & & \hspace*{3cm}
        \left.+Z_{(1,1)}\left( Z^0_0(\tau)^{16}-Z^0_1(\tau)^{16}-Z^1_0(\tau)^{16} \right)
              {Z^0_0(\tau)^*}^4 \right\} \ . \nonumber \\
  \label{SO(32)}
\end{eqnarray}
Note that the contribution from $X$ diverges when $q=\frac1R$ since the translational 
invariance in 8-9 plane is restored. 
So we have replaced $|Z^0_0(\tau)|^{-2}$ with the ordinary eta function. 

This partition function is the same as that for an orbifold of $SO(32)$ theory. 
To see this, rewrite $Z_{(\epsilon,\delta)}$ as follows. 
\begin{eqnarray}
Z_{(\epsilon,\delta)} &=& \frac{\sqrt{\alpha'\tau_2}}{2R}\sum_{k,l\in {\bf Z}}
      \exp\left[-\pi\tau_2\left(\frac{\alpha'}{4R^2}(2k)^2
      +\frac{4R^2}{\alpha'}\left(l+\frac{\delta}2\right)^2\right)
      +2\pi i\tau_1(2k)\left(l+\frac{\delta}2\right)
      \right] \nonumber \\
  & & \hspace*{-1.5cm}+(-1)^\epsilon\frac{\sqrt{\alpha'\tau_2}}{2R}\sum_{k,l\in {\bf Z}}
      \exp\left[-\pi\tau_2\left(\frac{\alpha'}{4R^2}(2k+1)^2
      +\frac{4R^2}{\alpha'}\left(l+\frac{\delta}2\right)^2\right)
      +2\pi i\tau_1(2k+1)\left(l+\frac{\delta}2\right)
      \right]  \nonumber \\
  &\equiv& {\cal Z}_{(0,\delta)}+(-1)^\epsilon {\cal Z}_{(1,\delta)}
\end{eqnarray}
These are familiar summations for the momentum and winding modes, except that there are 
winding modes with half-integral winding numbers. 
Then the partition function (\ref{SO(32)}) takes the following form. 
\begin{eqnarray}
&& Z_{1/R}(1,0^{15}) \nonumber \\
&=& \int \frac{d^2\tau}{\tau_2}\tau_2^{-5}|\eta(\tau)|^{-16}
    \left[
     {\cal Z}_{(0,0)}
       \left(Z^0_0(\tau)^{16}+Z^0_1(\tau)^{16}-Z^1_0(\tau)^{16}\right){Z^1_0(\tau)^*}^4
       \right. \nonumber \\
& & \hspace*{3cm}
    -{\cal Z}_{(1,0)}
       \left(Z^0_0(\tau)^{16}+Z^0_1(\tau)^{16}-Z^1_0(\tau)^{16}\right){Z^1_0(\tau)^*}^4
        \nonumber \\
& & \hspace*{3cm}
    +{\cal Z}_{(0,1)}\left\{
       \left(Z^0_0(\tau)^{16}-Z^0_1(\tau)^{16}\right)
       \left({Z^0_0(\tau)^*}^4+{Z^1_0(\tau)^*}^4\right) \right.
        \nonumber \\
& &\hspace{5cm} \left. +Z^1_0(\tau)^{16}\left({Z^0_1(\tau)^*}^4-{Z^0_0(\tau)^*}^4\right)
       \right\}  \nonumber \\
& & \hspace*{3cm}
    +{\cal Z}_{(1,1)}\left\{
        \left(Z^0_0(\tau)^{16}-Z^0_1(\tau)^{16}\right)
        \left({Z^0_1(\tau)^*}^4-{Z^0_0(\tau)^*}^4\right) \right.
        \nonumber \\
& & \left.\left. \hspace*{5cm}
       +Z^1_0(\tau)^{16}\left({Z^0_1(\tau)^*}^4+{Z^0_0(\tau)^*}^4 \right)\right\} \right]
\end{eqnarray}
This corresponds to an orbifold studied in \cite{hetorb}. 
It can be shown that such orbifold in the $R\to 0$ limit becomes   
non-supersymmetric $SO(32)$ theory. 
In fact, by taking the limit, 
\begin{equation}
Z_{1/R}(1,0^{15}) \to \int \frac{d^2\tau}{\tau_2}\tau_2^{-5}|\eta(\tau)|^{-16} 
         \left( Z^0_0(\tau)^{16}{Z^0_0(\tau)^*}^4-Z^0_1(\tau)^{16}{Z^0_1(\tau)^*}^4
               -Z^1_0(\tau)^{16}{Z^1_0(\tau)^*}^4 \right)\ ,
\end{equation}
which is the partition function of non-supersymmetric $SO(32)$ theory \cite{Pol}. 

\vspace{2mm}

Similar result can be obtained in $E_8\times E_8$ theory. 
In this case, the partition function is 
\begin{eqnarray}
Z_{1/R}(1,0^{15}) 
   &=& \int \frac{d^2\tau}{\tau_2}\tau_2^{-5}|\eta(\tau)|^{-16}Z_{E_8}(\tau)
       \left\{ Z_{(1,0)}\left( Z^0_0(\tau)^{8}+Z^0_1(\tau)^{8}-Z^1_0(\tau)^{8} \right)
              {Z^1_0(\tau)^*}^4 \right. \nonumber \\
   & &  \hspace*{4cm}
              +Z_{(0,1)}\left( Z^0_0(\tau)^{8}-Z^0_1(\tau)^{8}+Z^1_0(\tau)^{8} \right)
              {Z^0_1(\tau)^*}^4 \nonumber \\
   & &  \hspace*{4cm}
       \left.+Z_{(1,1)}\left( Z^0_0(\tau)^{8}-Z^0_1(\tau)^{8}-Z^1_0(\tau)^{8} \right)
              {Z^0_0(\tau)^*}^4 \right\} \nonumber \\
   & & \hspace*{-2cm}\to
       \int \frac{d^2\tau}{\tau_2}\tau_2^{-5}|\eta(\tau)|^{-16}Z_{E_8}(\tau) 
         \left( Z^0_0(\tau)^{8}{Z^0_0(\tau)^*}^4-Z^0_1(\tau)^{8}{Z^0_1(\tau)^*}^4
               -Z^1_0(\tau)^{8}{Z^1_0(\tau)^*}^4 \right)\ , \nonumber \\
\end{eqnarray}
where 
\begin{equation}
Z_{E_8}(\tau) = Z^0_0(\tau)^8+Z^0_1(\tau)^8+Z^1_0(\tau)^8\ .
\end{equation}
Therefore the resulting theory is non-supersymmetric $E_8\times SO(16)$ theory.

\vspace{1cm}

(ii) Non-SUSY theory without tachyon

\vspace{2mm}

Non-supersymmetric $SO(16)\times SO(16)$ theory can also be interpreted as supersymmetric 
heterotic theory on the Melvin background. 
Consider first $E_8\times E_8$ theory. 
The partition function (\ref{PartFn}) is then
\begin{eqnarray}
&& Z_{1/R}(1,0^7,1,0^7) \nonumber \\
   &=& \int \frac{d^2\tau}{\tau_2}\tau_2^{-5}|\eta(\tau)|^{-16}
       \left\{ Z_{(1,0)}\left( Z^0_0(\tau)^8+Z^0_1(\tau)^8-Z^1_0(\tau)^8 \right)^2
              {Z^1_0(\tau)^*}^4 \right. \nonumber \\
   & & \hspace*{3cm} 
        +Z_{(0,1)}\left( Z^0_0(\tau)^8-Z^0_1(\tau)^8+Z^1_0(\tau)^8 \right)^2
              {Z^0_1(\tau)^*}^4 \nonumber \\
   & & \hspace*{3cm}
        \left.-Z_{(1,1)}\left( Z^0_0(\tau)^8-Z^0_1(\tau)^8-Z^1_0(\tau)^8 \right)^2
              {Z^0_0(\tau)^*}^4 \right\}  \nonumber \\
&\to& \int \frac{d^2\tau}{\tau_2}\tau_2^{-5}|\eta(\tau)|^{-16}
       \left\{ \left( Z^0_0(\tau)^8+Z^0_1(\tau)^8-Z^1_0(\tau)^8 \right)^2
              {Z^1_0(\tau)^*}^4 \right. \nonumber \\
   & & \hspace*{3cm} 
        +\left( Z^0_0(\tau)^8-Z^0_1(\tau)^8+Z^1_0(\tau)^8 \right)^2
              {Z^0_1(\tau)^*}^4 \nonumber \\
   & & \hspace*{3cm}
        \left.-\left( Z^0_0(\tau)^8-Z^0_1(\tau)^8-Z^1_0(\tau)^8 \right)^2
              {Z^0_0(\tau)^*}^4 \right\} \ ,
  \label{SO16^2}
\end{eqnarray}
in the $R\to 0$ limit. 
The contribution from fermions to the partition function (i.e. the terms inside the curly 
bracket) can be rewritten as follows. 
\begin{eqnarray}
&& \hspace*{5mm}
   2\left\{Z^0_0(\tau)^8-Z^0_1(\tau)^8\right\}Z^1_0(\tau)^8
    \left\{{Z^0_0(\tau)^*}^4+{Z^0_1(\tau)^*}^4\right\}
    \nonumber \\
&&+\left[\left\{Z^0_0(\tau)^8+Z^0_1(\tau)^8\right\}^2+Z^1_0(\tau)^{16}\right]
      \left\{{Z^0_0(\tau)^*}^4-{Z^0_1(\tau)^*}^4\right\} \nonumber \\
&&-\left[\left\{Z^0_0(\tau)^8-Z^0_1(\tau)^8\right\}^2+Z^1_0(\tau)^{16}\right]
      {Z^1_0(\tau)^*}^4
     \nonumber \\
&&-2\left\{Z^0_0(\tau)^8+Z^0_1(\tau)^8\right\}Z^1_0(\tau)^8{Z^1_0(\tau)^*}^4 
\end{eqnarray}
Therefore the partition function (\ref{SO16^2}) exactly coincides with that for 
non-supersymmetric $SO(16)\times SO(16)$ theory (see, e.g. \cite{hetorb}). 

The same partition function can be obtained from supersymmetric $SO(32)$ theory. 
\begin{eqnarray}
&& Z_{1/R}\left(\left(\frac12\right)^8,0^8\right) \nonumber \\
&\to& \int \frac{d^2\tau}{\tau_2}\tau_2^{-5}|\eta(\tau)|^{-16}
       \left\{ 2Z^0_0(\tau)^8Z^0_1(\tau)^8{Z^1_0(\tau)^*}^4 \right. \nonumber \\
 & &\hspace*{1cm}  \left.
        +2Z^0_0(\tau)^8Z^1_0(\tau)^8{Z^0_1(\tau)^*}^4 
        -2Z^0_1(\tau)^8Z^1_0(\tau)^8{Z^0_0(\tau)^*}^4 \right\} 
\end{eqnarray}
One can easily see that this is the same quantity as (\ref{SO16^2}). 

It seems interesting that there exist such dual descriptions for $SO(16)\times SO(16)$ 
theory which has no tachyon in the spectrum. 
In the IIA-0A case, it is argued that the instability of the Melvin background in IIA 
picture is identified with the one due to the presence of tachyon in 0A picture 
\cite{Melvin}\cite{fluxbrane}. 
So it is interesting to identify the instability due to the Melvin background in terms of 
$SO(16)\times SO(16)$ theory. 
It is known that although $SO(16)\times SO(16)$ theory does not have tachyon at all, the 
perturbative vacuum of this theory is unstable due to a quantum effect \cite{nonhet}. 
Thus the dual descriptions might enable one to understand such instability.

\vspace{1cm}

(iii) Supersymmetric theory

\vspace{2mm}

The final example is supersymmetric theory on the Melvin background without Wilson line. 
Then the contribution to the partition function from the left-moving fermions is the ordinary 
one; $Z_{SO(32)}(\tau)$ or $Z_{E_8}(\tau)^2$, where
\begin{equation}
Z_{SO(32)}(\tau) = Z^0_0(\tau)^{16}+Z^0_1(\tau)^{16}+Z^1_0(\tau)^{16}\ .
\end{equation}
The partition function (\ref{PartFn}) in this case is 
\begin{equation}
Z_{1/R}(0^{16}) \to \int \frac{d^2\tau}{\tau_2}\tau_2^{-5}|\eta(\tau)|^{-16}Z_{left}(\tau)
       \left\{{Z^1_0(\tau)^*}^4+{Z^0_1(\tau)^*}^4-{Z^0_0(\tau)^*}^4\right\}\ ,
\end{equation}
in the $R\to 0$ limit. 
Therefore this suggests that a supersymmetric theory on the critical Melvin background is 
equivalent to the original supersymmetric theory. 
This is a new feature of this kind of duality, which does not occur in Type IIA case. 
The spacetime fermions emerge from the twisted sector, if the theory is interpreted as the 
orbifold $SO(32)$ or $E_8\times E_8$ on $S^1/(-1)^{F_s}\cdot \sigma^{\frac12}$.

\vspace{1cm}

\section{Discussions}

\vspace{5mm}

We have investigated the duality between supersymmetric heterotic theories on special 
backgrounds and nonsupersymmetric heterotic theories. 
This is based on the fact that the partition functions of two theories coincide. 
It is necessary to find further evidence which supports this duality conjecture. 
Fortunately, now we know some nonperturbative aspects of string theory. 
The duality web may be useful to study the properties of such backgrounds. 
For example, Type IIA-heterotic duality or Type I-heterotic duality will help us understand 
a mechanism which stabilizes the closed string tachyons. 
It will also be important to understand the instability in heterotic theories from the 
eleven-dimensional point of view, as has been done in \cite{Melvin}. 

In heterotic theories, the Melvin background corresponds to various perturbative vacua 
which are determined by Wilson line. 
However the stability property of the vacua is not common among them. 
The typical examples are; 
(i) unstable vacuum with tachyons, 
(ii) unstable vacuum without tachyon, 
(iii) stable vacuum. 
The instability of the vacua will be due to that of the Melvin background, so it is 
interesting to clarify how such a rich pattern can be realized by the presence of Wilson 
line. 

One of the important notion to understand the open string tachyon condensation is that this 
phenomenon can be described by a worldsheet theory. 
In particular, the interpretation of tachyon condensation by a RG flow induced by a boundary 
perturbation has provided an exact treatment \cite{BSFT} of this phenomenon. 
It will be very helpful that closed string tachyon condensation can also be understood in 
terms of a RG flow. 
Surprisingly, string theory on the Melvin background defines a family of CFT which 
interpolates between a theory with tachyon and another theory without tachyon. 
So there exists an exactly marginal deformation which connects the above two theories. 
Such a deformation will be related to the change in the flux, 
in terms of supersymmetric theory. 
Then the interpretation of the deformation in terms of non-supersymmetric theory will be 
important to understand the closed string tachyon. 

What we have discussed so far is a non-supersymmetric version of the target space duality. 
This resembles T-duality and Mirror symmetry in the sense that two different string theories 
on different backgrounds define the same theory. 
It is very interesting to know whether this is a special case of a more general duality, 
although the analysis will be difficult due to the absence of supersymmetry. 
There may be another examples of this kind of duality. 

In particular, it is interesting if there exists the following situation; 
there is a string theory which is dual to a background of another string 
theory, but the background is not connected with the trivial one in the space of solutions 
of the latter theory. 
Then it can be considered that the two theories are related by an off-shell deformation. 
In this situation, if exists, a nonperturbative effect will make a drastic change in the 
theory. 
It may force us to reconsider the connectivity of the moduli space of string theory. 

\newpage

\vspace{1cm}

{\Large {\bf Acknowledgements}}

\vspace{5mm}

I would like to thank S. Iso and H. Itoyama for valuable discussions. 

\vspace{1cm}

\appendix

\section{Modular property of partition functions}

\vspace{5mm}

In this appendix, we summarize the modular properties of various functions. 
The basic function is the theta function with characteristics. 
\begin{eqnarray}
\vartheta{a \brack b}(\nu,\tau) 
  &=& \sum_{n \in {\bf Z}}\exp\left[\pi i(n+a)^2\tau+2\pi i(n+a)(\nu+b) \right] \nonumber \\
  &=& \exp\left[\pi ia^2\tau+2\pi ia(\nu+b)\right]\vartheta_3(\nu+a\tau+b|\tau) 
\end{eqnarray}
The modular property of this function with $\nu=0$ is as follows. 
\begin{eqnarray}
&& \vartheta{a \brack b}(0,\tau+1) = e^{-\pi ia(a+1)}\vartheta{a \brack a+b+1/2}(0,\tau) \\
&& \vartheta{a \brack b}\left(0,-\frac1{\tau}\right)
    = (-i\tau)^{\frac12}e^{2\pi iab}\vartheta{b \brack -a}(0,\tau) 
\end{eqnarray}
Note that this function is periodic in $a$ and $b$, up to phase. 
\begin{eqnarray}
&& \vartheta{a+1 \brack b}(0,\tau) = \vartheta{a \brack b}(0,\tau) \\
&& \vartheta{a \brack b+1}(0,\tau) = e^{2\pi ia}\vartheta{a \brack b}(0,\tau) 
\end{eqnarray}

Our main interest is the partition functions (\ref{PFuns}). 
Due to the phase factors, the modular property of the functions is so simple. 
\begin{equation}
  \begin{array}{ll}
            Z^{0,2\theta_1}_{0,2\theta_2}(\tau+1)
              = e^{-\frac{\pi}{12}i}Z^{0,2\theta_1}_{1,2(\theta_1+\theta_2)}(\tau), & 
            Z^{0,2\theta_1}_{1,2\theta_2}(\tau+1)
              = e^{-\frac{\pi}{12}i}Z^{0,2\theta_1}_{0,2(\theta_1+\theta_2)}(\tau) \\
                     &  \\
            Z^{1,2\theta_1}_{0,2\theta_2}(\tau+1)
              = e^{\frac{\pi}6i}Z^{1,2\theta_1}_{0,2(\theta_1+\theta_2)}(\tau), & 
            Z^{1,2\theta_1}_{1,2\theta_2}(\tau+1)
              = e^{\frac{\pi}6i}Z^{1,2\theta_1}_{1,2(\theta_1+\theta_2)}(\tau) 
  \end{array}
\end{equation}
\vspace{2mm}
\begin{equation}
  \begin{array}{ll}
            Z^{0,2\theta_1}_{0,2\theta_2}\left(-\frac1{\tau}\right)
              = Z^{0,2\theta_2}_{0,-2\theta_1}(\tau), & 
            Z^{0,2\theta_1}_{1,2\theta_2}\left(-\frac1{\tau}\right)
              = Z^{1,2\theta_2}_{0,-2\theta_1}(\tau) \\
                   &    \\
            Z^{1,2\theta_1}_{0,2\theta_2}\left(-\frac1{\tau}\right)
              = Z^{0,2\theta_2}_{1,-2\theta_1}(\tau), & 
            Z^{1,2\theta_1}_{1,2\theta_2}\left(-\frac1{\tau}\right)
              = e^{-\frac{\pi}2i}Z^{1,2\theta_2}_{1,-2\theta_1}(\tau) 
  \end{array}
\end{equation}
One can easily see that the partition function (\ref{PartFn}) is indeed modular invariant.

\end{document}